# Identification of radiopure tungsten for low background applications

J. Hakenmüller[1] and W. Maneschg[1]

[1]*Max-Planck-Institut für Kernphysik, Saupfercheckweg 1, 69117 Heidelberg, Germany*

*e-mail addresses:*
`janina.hakenmueller@mpi-hd.mpg.de, werner.maneschg@mpi-hd.mpg.de`

**Abstract**

In this article we explore the availability of radiopure tungsten and its potential as high density shield material for low background applications. For compact shield designs, conventionally, lead is used. Metallic tungsten and tungsten pseudo-alloys reach higher densities up to $19.3\,\mathrm{g\,cm^{-3}}$ and do not exhibit a significant $^{210}$Pb activity, which is a typical intrinsic contamination in lead. Within several $\gamma$-ray screening campaigns we were able to identify tungsten samples with activities similar or better than $1\,\mathrm{mBq\,kg^{-1}}$ in $^{232}$Th, $^{40}$K, $^{60}$Co and the second part of the $^{238}$U decay chain. In cooperation with a manufacturer we further reduced a persisting contamination in the first part of the $^{238}$U decay chain by a factor of $\sim$2.5 down to $(305\pm30)\,\mathrm{mBq\,kg^{-1}}$. With Monte Carlo simulations, the construction of prototype tungsten-based setups and dedicated measurements, the shield capability of tungsten in comparison to lead was extensively studied. Specifically, the impact of cosmogenic radiation on the background at shallow depth was investigated. We showed that a 6-40% reduction (depending on the exact shield configuration) in the muon-induced neutron fluence is achievable by replacing lead with an equivalent amount of tungsten regarding the suppression of external $\gamma$-radiation. Overall, many benefits using tungsten especially for low energy applications below a few 100 keV are found. The pseudo-tungsten alloy presented in this work paves the way for several applications especially regarding background suppression in particle and astroparticle physics search programs.

## 1 Introduction

Shields for low background applications in particle and astroparticle physics require materials, that have intrinsically very small concentrations of radioactive impurities. This radiopurity constraint limits the number of materials that can be considered at all, and repeatedly requires long-lasting radioassay campaigns before the construction of detectors and shields. Typically used structure support materials include radiopure carbon steel (CS; e.g. [1, 2, 3]), stainless steel (SST, e.g. [2, 4, 5, 6, 7, 8]) or titanium (Ti) (e.g. [7, 9, 10]). In comparison to them, copper (Cu) and lead (Pb) have a lower mechanical strength, but higher densities, higher atomic numbers and eventually less radioactive impurities, which make them ideal candidates for the construction of compact shield setups. Electroformed Cu has been found to be the radiopurest metal [11], however Cu can be easily activated via cosmic radiation and thus longer exposures above ground must be strictly avoided. Pb has a 30% higher density than Cu and does not suffer from significant cosmic activation, but its radiopurity is typically lower and it is often contaminated with the long-lived isotope $^{210}$Pb (half-life: 22.3 yr). Pb with very low $^{210}$Pb concentrations below $1\,\mathrm{Bq\,kg^{-1}}$ is either of archaeological origin (and therefore very limited in quantity) or it is commercially available but expensive. Within this work, we discuss metallic tungsten (W) (99.9%) and W-based pseudo-alloys with high W content ($\sim$97%) as novel non-corrosive, high density, low background materials, that can be considered as a future replacement for radiopure Pb and Cu, at least to some extent and always depending on the exact demands on the shield. The advantages of W-based materials are summarized in the following:

- density and atomic number: even though the atomic number of W is lower than that of Pb, i.e. 183.8 versus 207.2 a.u., metallic W and W pseudo-alloy samples presented in this work achieve densities of 18.5-19.3 $\mathrm{g\,cm^{-3}}$, which are 60-70% higher than the density of Pb with $11.3\,\mathrm{g\,cm^{-3}}$.

- By considering the proper $\gamma$-ray attenuation coefficients on the MeV scale, an almost double thick Pb shield layer is needed to achieve the same shielding power as for W. This allows to build more compact shields with W, but it might be also beneficial in terms of cosmic muon suppression as will be shown within this article.

- melting point: while Pb has a very low melting point at 328°C, W has the highest melting point of all known metals of 3387-3422°C (variation in literature values). Further, W has the best high-temperature mechanical properties and the lowest expansion coefficient of all metals. A shield design comprising flammable materials with a surrounding W layer reduces the fire-load and enhances the safety e.g. in underground laboratories or at nuclear power plants.

- mechanical strength: while Pb is rather soft and can be deformed by a non-proper handling, W has an extreme hardness, i.e. 1.5 vs. 7.5 on the Moh's scale. Thus, well-shaped W shield components with plain surfaces can be produced. This is advantageous, because it helps reducing the size of potential gaps between different components and thus provides a better shield against external $\gamma$-radiation and penetrating air-borne radioactive $^{222}$Rn (half-life: 3.8 d). Further, machining plates is harder for W than for Pb due to the high W hardness, but on the other hand Pb easily smears out due to its extreme softness, which pollutes the machinery.

- contamination by natural radioactive U decay series: The inert U daughter isotope $^{222}$Rn escapes the Earth's crust and emanates into the atmosphere. $^{222}$Rn and its decay products like $^{210}$Po and $^{210}$Pb can attach to airborne particulates, which again can be washed out by rainfalls and be sedimented on Earth's surface [12]. Especially Pb productions suffer under both washout $^{210}$Pb sedimentation and air containing $^{210}$Pb aerosols. Once $^{210}$Pb is contained in the Pb, is cannot be separated by standard techniques due to their equal chemical properties. In the case of W and other non-Pb materials, elevated $^{210}$Pb contaminations are not observed. In general, $^{210}$Pb is of concern since it decays into $^{210}$Bi, which is a $\beta$-emitter with energies up to $\sim$1.2 MeV. This is critical especially for low energy applications below a few 100 keV.

- self-absorption: materials with a high-Z number (like W and Pb) are always advantageous to suppress the background continuum especially for the energy spectrum below a few 100 keV. This is due to the fact that the (e.g. muon-induced) *bremsstrahlung* ($\alpha\, Z^2$) continuum has a lower intensity due to a stronger self-absorption via the predominant photo-effect ($\alpha\, Z^{4-5}$) compared to materials such as Cu or steel. Further, both W and Pb have characteristic X-rays with comparable intensities at almost the same energy intervals, i.e. the K shell X-rays around 59-85 keV and the L shell X-rays around 8-15 keV. In these respects, both Pb and W behave similarly well and are therefore of great interest for innermost shield layers in low energy applications.

- hazard: Pb vapor or Pb dust is hazardous via incorporation. Pb residues on skin, clothes and equipment can be dispersed in homes after working with Pb at a workplace [13, 14]. Counter measures have to be undertaken to avoid this and non-compliance to safety rules can be dangerous. On the contrary, W is a non-toxic material, which is therefore – despite the usual higher costs in trade compared to Pb – an alternative to Pb not only for low background application in basic physics research, but as well in general regarding commercial nuclear and medical equipment [15].

Herein, we report about the search for radiopure high density W that is suitable as shield material for low background applications. Next to an introduction to the W production (Section 2), we present our results from dedicated W radioassay measurement campaigns and W production supervised by us (Section 3). With a series of Monte Carlo (MC) simulations and prototype shield setups (Section 4) we demonstrate that the achieved radiopurity levels in W can already improve Pb-based shield designs. Finally, Section 5 includes a summary and conclusions.

## 2 Tungsten production

Preliminary radioassay measurements of high density W samples from different manufacturers at the beginning of this project (cf. Section 3) confirmed that W samples are in general radioactively contaminated. The only positive exception were samples from the company H.C. Starck GmbH in Hermsdorf, Germany (HCS Hermsdorf). The company is one of six end-user material production sites located in China, Germany, Great Britain and USA [16], which belong to the Vietnamese Masan Tungsten Limited Liability Company (MTC) [17], a wholly-owned subsidiary of Masan Resources (MSR). Within a scientific cooperation, HCS Hermsdorf agreed to investigate with us the origin of the remaining contaminants occurring during the on-site W production. This section and Figure 1 illustrate

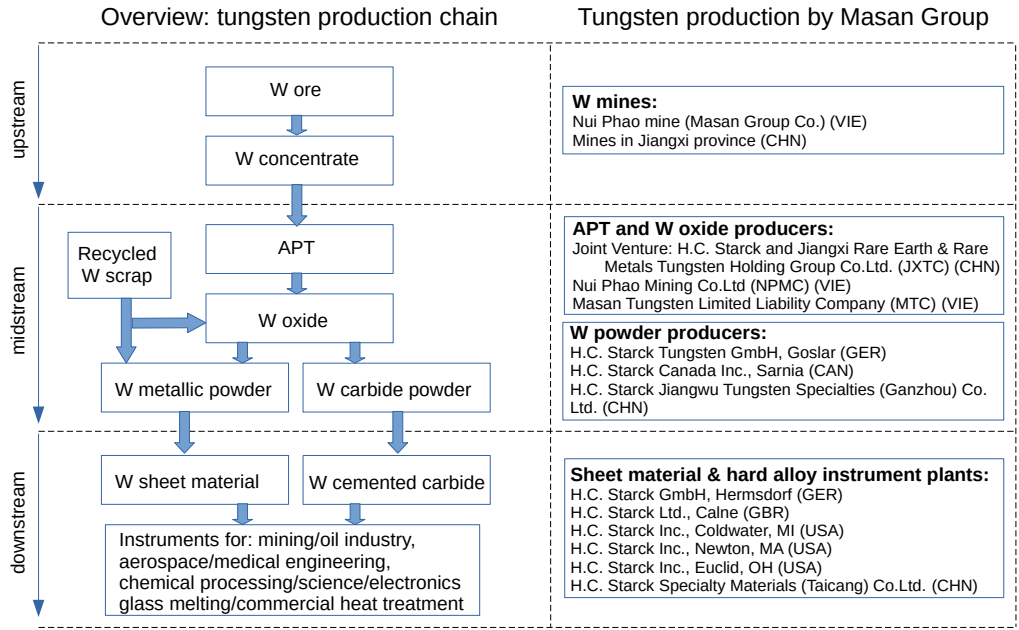

**Figure 1**: W value chain: general flow diagram and the specific case of the Masan Group Co. and H.C. Starck GmbH in Hermsdorf, Germany, who provided mainly the W samples for this work.

the up- to downstream value chain of W in the general and in the specific case of HCS Hermsdorf.

W is a rare metal, which is mostly found in the Earth's crust in combination with other elements or chemical compounds. The principal ores contain wolframite (black ore, chemically iron-manganese tungstate ((Fe,Mn)$WO_4$)) or scheelite (white-ore, chemically calcium tungstate ($CaWO_4$)) [18, 19]. The worldwide annual W mine production in 2020 amounted to 84,000 metric tons, wherein 82.1%, 5.1%, 2.6%, and 2.3% are attributed to the main producers China, Vietnam, Russia and Mongolia, respectively [20]. Beside this primary raw material supply, W scrap material contributes to ∼35% of the W supply chain [21]. H.C. Starck sources W mainly from the Nui Phao mine in Vietnam, but also from mines in the Jiangxi province in China and from scrap W material exclusively affiliated to the company itself.

In the first step, the ore is mechanically crashed, milled/grinded and the sub-components are separated (gravimetrically or magnetically) to obtain a concentrate with typically 65-70% of tungsten-oxide ($WO_3$). Second, the concentrate is dissolved by reactant alkaline solutions (e.g. a soda or a concentrated NaOH solution), purified and filtrated (e.g. via solvent extraction or ion-exchange resin), before being transformed into an ammonium tungstate solution (ATS). The latter is converted into ammonium para-tungstate (APT) via crystallisation containing at least 88.5% $WO_3$. APT is the most common precursor for many W products, since a thermal and chemical decomposition can be carried out easily. In a third step, APT is calcinated forming tungsten-oxide, which has 99.9% $WO_3$. Tungsten-oxide is reduced under a hydrogen atmosphere to obtain W metal powder. In the case of HCS Hermsdorf, this step is done at one of three H.C. Starck W powders facilities [22] around the world (Germany, Canada and China), namely at H.C. Starck Tungsten GmbH in Goslar, Germany.

In a last step, W metal powder is used either to produce W carbide powder (via carburization) for cemented carbide, or to obtain directly ferrotungsten, super-alloys and pseudo-alloys for high density hard metal/alloy work pieces.

For the production of the high density W sheets tested in this work, HCS Hermsdorf mixed W powders of selected types and densities of 3-4 g cm$^{-3}$ with small amounts of additives at the percent level. Additives can be Cu, Fe, nickel (Ni) or cobald (Co); in our case Fe (0.9%) and Ni (2.1%) were used exclusively. Next, the W compound was compressed isostatically via water compression at a pressure of the order of 100 MPa. At this stage, the density of the still porous W compound achieves ∼10 g cm$^{-3}$. The W compound was deposited into a fire-resistant transport tub, in which a thin film of clay ($Al_2O_3$) acts as separation medium. The tub was then moved through a rotary furnace at ∼1400°C under hydrogen atmosphere, dur-ing which the W compound was sintered. Sintering is a non-liquid powder metallurgy technique which is economically more convenient than melting, since W has a very high melting point (cf. Section 1) requiring a large energy consumption. During sintering the microscopically irregular-shaped W granulate gets a rounder shape and the additives fill completely the air gaps in-between. This elongated grain structure enhances the binding of the composite and leads to a stable, high density W pseudo-alloy. The W content and density of the final product called 'HPM1850' by HCS Hermsdorf is 97% and $18.5\,\mathrm{g\,cm^{-3}}$, respectively.

## 3 Tungsten radioassay campaigns

### 3.1 Exploration of the tungsten market

In the past, several authors searched for radiopure low-density W powders as well as high density metallic W and W pseudo-alloys. Different levels of radiopurity were found in $WO_3$ powders used e.g. for the production of $ZnWO_4$ [23], $CaWO_4$ [24] and $CdWO_4$ [25], which are crystal scintillators needed in double beta decay and dark matter search. The segregation process involved in the crystallation process often helped to achieve a higher radiopurity grade. In the case of high density metallic W and W pseudo-alloys, only samples elevated in $^{238}$U, $^{232}$Th and $^{40}$K activities of the order of 10-100 $\mathrm{mBq\,kg^{-1}}$ have been found so far [26].

To start our search for radiopure high density W composites, we looked systematically for available W samples on the global market. Finally, W plates from several manufacturers designed for the application within the experimental stellarator facility Wendelstein 7-X [27] became available to us for screening. All these plates had the same geometry and a mass of 4.4-4.6 kg.

The radiopurity of the samples was evaluated by means of $\gamma$-ray spectroscopy using the high purity Ge (HPGe) detectors CORRADO [5] and GIOVE [28] at the MPIK underground laboratory. It is located at shallow depth and has an overburden of 15 m of water equivalent (m w.e.). Typically, the samples were first cleaned with isopropanol and then measured with the material screening detector CORRADO. Its sensitivity for U and Th lies in the range of $1\,\mathrm{mBq\,kg^{-1}}$. For samples with minor contaminations, the measurement was repeated with the GIOVE detector. It achieves sensitivities in U and Th of $\sim 100\,\mathrm{\mu Bq\,kg^{-1}}$ for large sized samples with a total mass of several 10 kg.

Our results are summarized in Table 1. Distinctive differences were observed between the samples from different manufacturers. We confirmed elevated U, Th and K contamination levels in all cases except for the W pseudo-alloy from the company HCS Hermsdorf. In this case upper limits for most of the radioisotopes were deduced. Especially for $^{40}$K a superior performance was found. Only a contamination from the first part of the $^{238}$U decay chain – indicated by the presence of the 1001 keV line of $^{234m}$Pa – was observed. In this case the activity results to be $(761\pm67)\,\mathrm{mBq\,kg^{-1}}$. For the second part of the U decay chain following $^{230}$Th ($T_{1/2}$=7.5·10$^4$ yr) the activity was found to be small in comparison at a level of $(1.5\pm0.6)\,\mathrm{mBq\,kg^{-1}}$. This means that the secular equilibrium of the decay chain is broken. The inequality of the activities might origin from a removal of the second half of the decay chain during the manufacturing process or reciprocally from an additional introduction of $^{234m}$Pa e.g. within the furnace.

### 3.2 Investigation of the tungsten production chain at HCS Hermsdorf

In the consecutively established cooperation with HCS Hermsdorf it was possible to screen W powder samples as well high density W samples, that were produced both recently and $\sim$30 y ago. This includes APT as raw material (89.1% $WO_3$), various W powders of different grain sizes and densities as well as the Fe and Ni powders that are used as additives in the production of here discussed W pseudo-alloys (cf. Section 2). Bottles with various powders provided by the manufacturer are displayed in Figure 2. Further, components belonging to the sintering furnace were examined.

Measurements of the adhesive Ni and Fe powders demonstrated that they do not contribute to the observed $^{238}$U contamination of the final material. Measurements of the APT powder showed only an upper limit of $<250\,\mathrm{mBq\,kg^{-1}}$ for $^{238}$U. The same holds for some of the powders. Most notably, however, the powder denoted 'D80' turned out to be contaminated in the first part of the $^{238}$U decay chain.

Next, new and already used $Al_2O_3$ powders needed to keep the W compound separated from the underlying transport tub when moving through the sintering furnace (cf. Section 2) were measured. Nearly no $^{234m}$Pa decays are observed in the new material, the used material, however, shows a slightly stronger contamination. $^{40}$K is present in both samples, see Figure 3.

Additionally, other materials in close proximity with the W compound during the sintering process were examined. These include the fire resistant stone linings covering the inner surface of the rotary furnace. These stones are made out of fire clay with a large share of $Al_2O_3$ to be able to withstand the high temperature. Pieces of new furnace stones as well as used ones with visible carbon black pollution were investigated with the CORRADO detector. The new furnace stones exhibit contaminations from the U and Th decay chains and from $^{40}$K, while the used ones show the same contaminants but amplified by almost two orders of magnitude, see Figure 4.

**Table 1**: Comparison of HPGe spectroscopy measurements of W samples of the same geometry provided by various manufacturers. All activities are reported in units of $\mathrm{mBq\,kg^{-1}}$. The measurements were carried out by default with the CORRADO detector. Samples that showed a low activity were remeasured with the more sensitive GIOVE detector.

| manufacturer | Plansee | Plansee | Osnabruegge | MG Sanders | HCS Hermsdorf |
|---|---|---|---|---|---|
| composition | 100% W | pseudo-alloy | 100% W | 100% W | 97% W pseudo-alloy |
| mass kg$^{-1}$ | 4.51 | 4.36 | 4.54 kg | 4.60 kg | 4.50 |
| density/g/cm$^3$ | 19.3 | 18.5 | 19.3 | 19.3 | 18.5 |
| cleaning | isopropanol | isopropanol | isopropanol | isopropanol | isopropanol |
| measurement time/d | 5.0 | 16.2 | 4.5 | 4.1 | 18.8 |
| detector | CORRADO | CORRADO | CORRADO | CORRADO | GIOVE |
| $^{238}$U | <417 | 468±132 | <1028 | 631±299 | 761±67 |
| $^{226}$Ra | 150±9 | 59±4 | 1075±36 | 29±7 | 1.5±0.6 |
| $^{228}$Th | 55±7 | 22±3 | 167±14 | 469±26 | <0.7 |
| $^{228}$Ra | 54±7 | 26±4 | 246±17 | 349±20 | <1.2 |
| $^{40}$K | <41 | 33±11 | <73 | 38±20 | <5.7 |
| $^{60}$Co | <1.9 | <2.0 | <3.2 | <2.2 | <0.3 |
| $^{137}$Cs | <2.5 | <1.9 | <4.5 | <5.1 | <0.6 |

![Assembly of W powders bottles]

**Figure 2**: Assembly of all W powders provided by HCS Hermsdorf for HPGe $\gamma$-ray spectroscopic measurements.

Finally, we investigated 30 y old metallic W plates from HCS Hermsdorf and confirmed that the $^{234m}$Pa activity is already present there. This excludes the possibility that a recent change in the manufacturing process has introduced the observed $^{234m}$Pa contamination.

To summarize, we identified the powder 'D80' as the only one material with a significant $^{234m}$Pa contamination, while different components in the manufacturing process seem to accumulate different radioactive traces over time.

![HPGe gamma-ray spectroscopy plot of Al2O3 powders]

**Figure 3**: HPGe $\gamma$-ray spectroscopy of Al$_2$O$_3$ powders: the red spectrum corresponds to new material, while the black spectrum represents 1 yr old material continuously used in W productions.

### 3.3 Dedicated W production at HCS Hermsdorf under radiopurity constraints

Aground the previous findings in Section 3.2, we ordered W plates from HCS Hermsdorf in a two-step approach and supervised the production to further increase the radiopurity level. This excluded the usage of the 'D80' powder and included the usage of new/old furnace stones, carrying gloves during the entire manufacturing process and the application of grinding and ultrasonic bath for the final W plates.

Batch #1 (4 plates, each 10.4 kg) was produced with new furnace stones, while batch #2 (28 plates, each 10.4 kg) was produced using old furnace stones. Figure 5 displays 10 plates from batch #2 after deployment into the sample chamber of the GIOVE detector. Table 2 summarizes our findings. The contamination in $^{238}$U is still present in batch #1 and #2, but was significantly reduced by factor ∼7 compared to the initial product. Batch #2 also revealed a small contribution from the $^{232}$Th decay chain.

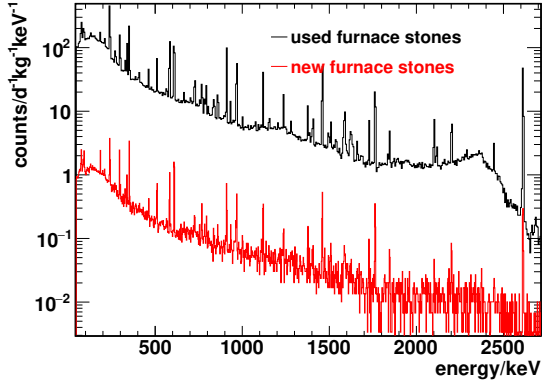

**Figure 4**: HPGe γ-ray spectroscopy of furnace stones from W production in a new state (red spectrum) and in a used state (black spectrum) after having served as furnace lining for ∼1 yr.

**Table 2**: HPGe γ-ray spectroscopy of W samples from HCS Hermsdorf. The activities are given in units of mBq kg$^{-1}$. The W plates were produced in two batches by excluding the powder type 'D80' and by strictly avoiding any accidental radioactive contaminations.

| sample | 4 plates | 10 plates |
|---|---|---|
| total mass/kg | 41.6 | 104.0 |
| density/(g cm$^{-3}$) | 18.5 | 18.5 |
| cleaning | isopropanol | isopropanol |
| measurement time/d | 15.7 | 13.9 |
| screening detector | GIOVE | GIOVE |
| $^{238}$U | 284±26 | 305±30 |
| $^{226}$Ra | 7.9±0.4 | 4.2±0.3 |
| $^{228}$Th | 0.7±0.2 | 1.0±0.2 |
| $^{228}$Ra | 1.9±0.4 | 1.3±0.4 |
| $^{40}$K | <1.2 | 5.0±0.9 |
| $^{60}$Co | <0.06 | <0.08 |
| $^{137}$Cs | <0.2 | <0.3 |

To conclude, we successfully produced W plates with a high radiopurity level and used them in the construction and testing of a prototype shield made of W and Pb. The outcome of these further investigations is described in Section 4.

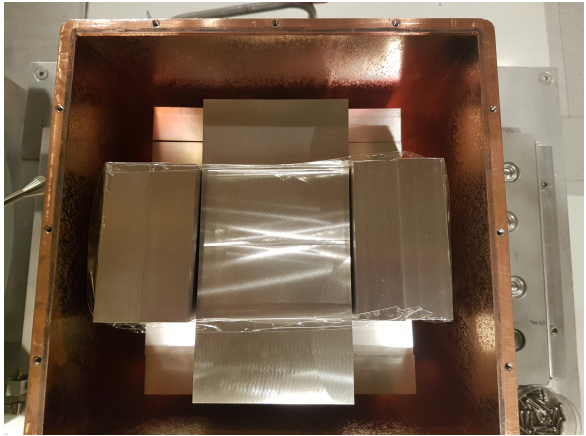

**Figure 5**: Sample of 10 W plates from batch #2 deployed into the GIOVE sample chamber. The dimensions of a single plate are: 20 cm×10 cm×2.8 cm.

### 3.4 Investigation of the cosmogenic activation in W

One major problem for low background materials might be their enhanced probability to accumulate radioactive isotopes with long half-lives, when being exposed to proton or neutron radiation. This can occur e.g. via spallation reactions induced by high energetic radiation or via neutron capture.

In the case of artificial highly energetic irradiation of W targets e.g. at the European Spallation Source or the Joint European Torus tokamak, the spallation processes have already been thoroughly studied [29, 30]. The impact of cosmic rays at Earth's surface on W samples was previously investigated within the dark matter experiment CRESST, which uses CaWO$_4$ crystals as detectors. The authors focussed on the energy range below 80 keV and found two activation processes involving tantalum (Ta) and hafnium (Hf) isotopes [31]. First, $^{182}_{74}$W(p,α)$^{179}_{73}$Ta is followed by $^{179}_{73}$Ta(e$^-$,ν$_e$)$^{179}_{72}$Hf$^*$, wherein $^{179}_{72}$Hf$^*$ deexcites into $^{179}_{72}$Hf plus X-rays with energies at 2.60, 10.74, 11.27 and 65.35 keV. Second, $^{183}_{74}$W(p,$^3$He)$^{181}_{74}$W is followed by $^{181}_{74}$W(e$^-$,ν$_e$)$^{181}_{73}$Ta$^*$, wherein $^{181}_{73}$Ta$^*$ decays under emission of γ- and X-rays at 6.2 and 67.4 keV. Within this work we explored the energy spectrum of our high density W samples (cf. Table 2) from 80 keV up to 3 MeV. We found a series of γ-lines, that might be attributed to cosmic activation. The corresponding energies, count rates and some radioisotope candidates from Ta, Hf and lutetium (Lu) are listed in Table 3. In one single case the radioisotope was unambiguously identified: $^{182}_{73}$Ta. It has a half-life of 114.6 d, a Q-value of 1814 keV and it can be produced in the reactions $^{182}_{74}$W(n,p)$^{182}_{73}$Ta, $^{183}_{74}$W(n,p+n)$^{182}_{73}$Ta, $^{183}_{74}$W(n,d)$^{182}_{73}$Ta, $^{184}_{74}$W(n,d+n)$^{182}_{73}$Ta and $^{186}_{74}$W(n,d+3n)$^{182}_{73}$Ta. By knowing the cosmogenic activation history, the γ-ray emission probabilities and the simulated detector-sample efficiency, it was possible to determine a sea level production rate of (121±30) atoms d$^{-1}$ for this isotope.

Additionally, neutron capture reactions from

7**Table 3**: Gamma-lines of confirmed/potential cosmogenic origin found in the newly produced W samples by HCS Hermsdorf. The count rates were obtained from the 104 kg W sample shown in Figure 5 and deployed inside the sample chamber of the GIOVE detector. Isotope candidates that could not be fully identified in terms of branching ratios and half-lives are set in brackets. The literature values are taken from [33, 34].

| peak position / keV | | count rate | isotope |
|---|---|---|---|
| fit | literature | / (counts d$^{-1}$) | |
| 1231.2±0.3 | 1231.004(3) | 1.7±0.5 | $^{182}_{73}$Ta |
| 1221.0±0.2 | 1221.395(3) | 2.6±0.5 | $^{182}_{73}$Ta |
| 1188.7±0.4 | 1189.040(3) | 1.6±0.4 | $^{182}_{73}$Ta |
| 1094.2±0.6 | 1093.657(13) | 0.9±0.3 | ($^{172}_{71}$Lu) |
| 944.9±0.3 | 942.80(11) | 2.3±0.6 | ($^{178m}_{72}$Hf) |
| 898.3±0.5 | 900.724(20) | 1.3±0.6 | ($^{172}_{71}$Lu) |
| 572.7±0.5 | 574.215(21) | 3.1±1.7 | ($^{178m2}_{72}$Hf) |
| 534.9±0.4 | 535.036(18) | 2.4±0.7 | ($^{178m2}_{72}$Hf) |
| 110.9±0.5 | 110.39; 113.67 | 2.3±1.0 | $^{182}_{73}$Ta |

mostly thermal neutrons can occur from cosmic rays at the Earth's surface, but also within a radiation shield. Considering the natural abundance of the different W isotopes and the fact that the capture increases the neutron number of an isotope by one, the instable isotopes created in $^{184}$W(n,γ)$^{185}$W (half-life of $^{185}$W: 75.1 d) and in $^{186}$W(n,γ)$^{187}$W (half-life of $^{187}$W: 23.72 h) are identified as potential candidates. For the first reaction the thermal neutron capture cross section is 2 b, while for the latter it is 50 b [32]. $^{185}$W predominately undergoes β-decay. Within this work the lines at 686.6 keV and 479.5 keV from the decay of the short-lived $^{187}$W from activation at the Earth's surface were observed at the start of the screening measurements. The contribution of the line count rates from *in situ* activation by the neutron fluence within the shield was found to be below the sensitivity of the measurement. Assuming a neutron fluence rate of 20 cm$^{-2}$s$^{-1}$ within the GIOVE shield, for a W sample size of 104 kg and the sample placed directly at the detector for the most prominent line of $^{187}$W at 686.6 keV less than 0.03 counts d$^{-1}$ are expected. This result includes the geometric detection efficiency derived from MC.

## 4 Tungsten based shield designs: MC simulations and prototypes

To understand the impact of W as a new high density material in low background applications in comparison with specifically Pb, but also Cu and Fe, we simulated simple and more elaborated shield designs in regard of suppression of external γ-radiation, cosmic ray background at shallow depth (<100 m w.e.) and intrinsic radioactive contaminations, which include our experimental W screening results as normalization. Moreover, we built a prototype W-Pb hybrid shield integrating the newly acquired W plates. As detection medium in both the MC simulation and the experimental approach we selected our mobile HPGe spectrometer CONRAD, which consists of a ∼2 kg massive Ge diode inside a low background copper cryostat [35]. The MC simulations were carried out within the Geant4-based simulation framework MaGe [36]. Specifically we used the version Geant4.10.3 [37, 38, 39], that we had validated beforehand in terms of electromagnetic, muon and neutron interactions [40]. In the simulation output, all energy depositions inside the Ge diode were registered as well as the energy spectrum and the number of neutrons entering through the diode surface.

To start we simulated two simple 4π shields made out of either pure W or pure Pb. A sketch of them is shown in Figure 6. The outcome is reported in Sections 4.1 to 4.3.

Finally, we simulated a more complex shield design by starting from the existing one of the GIOVE detector. Our results are presented in Section 4.4. The outcome of measurements with the prototype Pb-W hybrid are listed in Section 4.5.

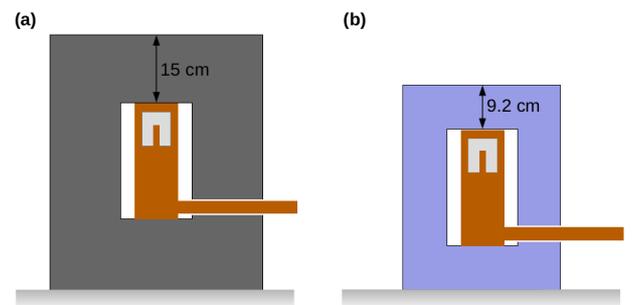

**Figure 6**: Simulation of a simple high density shield geometry surrounding a HPGe detector inside a copper cryostat and with a minimum thickness of (a) 15 cm Pb and (b) 9.2 cm W in all directions. The Pb shield has a mass of 1200 kg, while the equivalent in W has a mass of 810 kg.

### 4.1 Suppression of external γ-radiation

All low background applications including experiments looking for rare event rates require a proper shield against external γ-radiation. The major contribution comes from ambient natural radioactivity, but in specific cases other sources can also play an important role, e.g. radioactive decays of $^{16}$N isotopes in cooling cycles of nuclear power plants. The γ-radiation intensity *I* decreases exponentially when passing through matter with the suppression strength depending on the material thickness $x$, the density $\rho$ and the material-specific $\gamma$-ray attenuation coefficient $a$ (cf. tabulated values in [41]). The latter is related to the cross section of the underlying processes responsible for the attenuation:

$$I(x) = I(0) \cdot exp(-\rho \cdot a \cdot x) \quad (1)$$

These parameters need to be considered in the shield design. Materials with a high-Z number have a high density and large attenuation coefficients for $\gamma$-rays up to several MeV. In general, the coefficients decrease strongly towards higher energies. For W and Pb, they have minima around 3 MeV – the upper end of the $\gamma$-ray energies observed in natural radioactivity – and they increase again above due to the cross section for pair production increasing with energy.

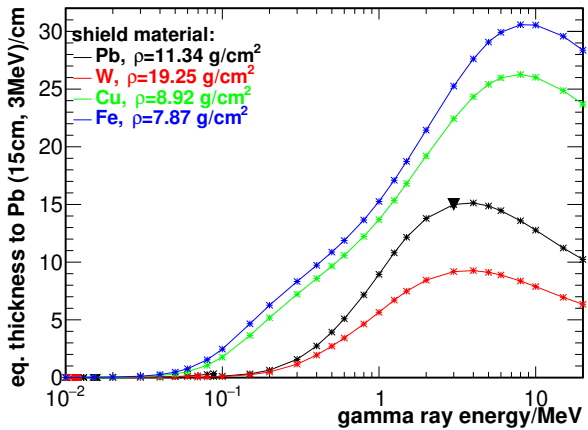

**Figure 7**: Equivalent thickness of shields made purely of Pb, W, Cu or Fe for energies in the range from 10 keV to 20 MeV in order to achieve the same background suppression against $\gamma$-radiation as from 15 cm Pb for 3 MeV $\gamma$-rays.

Within a realistic design as shown in Figure 6, we start from a 15 cm thick Pb shield and 3 MeV $\gamma$-rays. According to Eq. 1 one expects an attenuation of $7 \cdot 10^{-4}$ inside the Pb shield. To achieve the same $\gamma$-ray attenuation with W, Cu and Fe, shield thicknesses of correspondingly 9.2, 23 and 25 cm are required, see Figure 7. The superiority of W over Pb for the suppression of all $\gamma$-ray energies is mainly due to the significantly larger density, while the attenuation coefficient of W is indeed a few % smaller than that of Pb. For all materials, the required shield thickness decreases towards lower $\gamma$-ray energies, however, the differences between W and Pb become smaller as well. This complication needs to be taken into account when replacing Pb with W in a real setup. To see the consequences for the attenuation of 2615 keV $\gamma$-rays that mostly looses energy step-wise loose when crossing the shield, a mono-energetic isotropic point source outside the Pb/W shield in a distance of about 30 cm to the HPGe diodes was simulated. In the full spectral range of [40,2700] keV the suppression factor is almost identical for Pb and W. For W it is $1.70 \cdot 10^{-5}$, while for Pb it is slightly more effective by 5%. The expected background spectral shapes are comparable over the entire energy range.

### 4.2 Suppression of muon-induced background

A drawback of high-Z materials is that they provide a target for muons to produce secondary radiation. This is especially problematic at shallow depth, where the muon flux is not considerably suppressed. In this case, the background inside the shield can be dominated by the two main muon-induced contributions. The first component consists of prompt muon-induced electromagnetic radiation. It covers the full energy range relevant to HPGe $\gamma$-ray spectroscopy. Below $\sim$500 keV it leads to a characteristic *bremsstrahlung* continuum (see Figure 8). Muon-induced neutrons, the second component, contribute to the HPGe spectrum via elastic and inelastic scattering as well neutron capture. Especially at energies below 100 keV$_{ee}$, the neutron-induced recoils of Ge nuclei become highly relevant and result in a strong increase in events towards lower energies as shown in Figure 8. Energy depositions following nuclear recoils result in ionization energy and heat. HPGe spectrometers operated at liquid nitrogen temperature can only read out the ionizated charges, but not the phonons. This loss mechanism called 'quenching effect' needs to be considered, also in the MC simulations. For the course of this publication, we rely on the recent quenching result from [42].

We investigated the muon-induced background in HPGe detectors embedded inside W- and Pb-based shields assuming an overburden of 15 m w.e. The expected muon flux and spectrum at this depth have already been calculated and validated in our previous work. The same holds for the muon-induced spectra expected inside a shield at the position of the HPGe detector [40]. We simulated Pb shields with 15 and 25 cm thickness and their respective equivalent in W. The mass of the W shield is reduced by 30% or respective 40% (for a small detector chamber of size of 15 cm$\times$15 cm$\times$24 cm). The resulting count rates seen by the HPGe diodes from all muon-induced contributions and the respective muon-induced neutron fluence rates are summarized in Table 4.

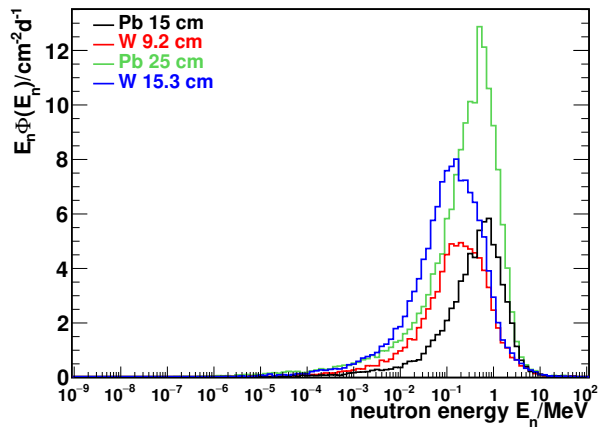

**Figure 8**: MC simulated muon-induced neutron fluence rate at the HPGe diode surface for Pb shields of 15 and 25 cm and for the respective equivalent of W. It becomes evident, that within the W shield about 17% (27%) less fast neutrons are prevalent.

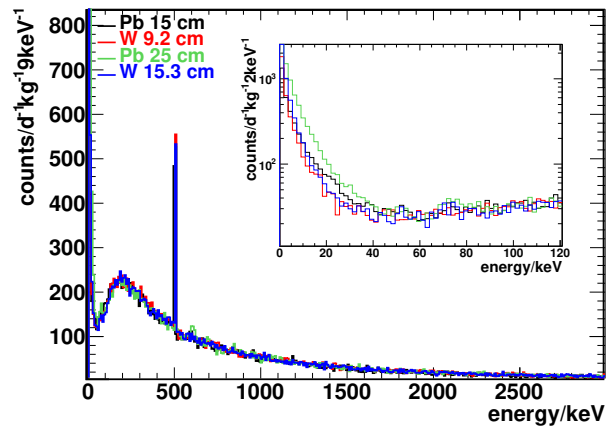

**Figure 9**: MC simulated background spectra from muon-induced secondaries that are created inside Pb shields of 15 cm and 25 cm and inside the respective equivalent of W. The impact of the lower neutron fluence rate inside the W shield can be observed in the reduced continuum below $\sim$60 keV.

For the two Pb setups, the electromagnetic contributions agree above 100 keV within the uncertainties, as the self-shielding effect of Pb successfully suppresses all additional contributions from the outer layers of the larger shield, see Figure 9. However, at lower energies, where the neutron recoils dominate, the additional contributions by the larger shield become apparent. According to our simulations, the neutron fluence rate at the diode surface is namely doubled, which leads to the respective enhancement in the HPGe spectrum.

Replacing Pb with W results in a significant reduction in the number of fast neutrons by 17% for the smaller shield and by 26% for the larger one. A notably shift in the neutron spectrum towards lower energies can also be observed in Figure 8 in the case of the W shield. This facilitates the moderation of these neutrons towards thermal energies. The reduced production of fast neutrons translates directly into a reduction of the count rate below 100 keV by at least 20%, see Figure 9. The only exception is the count rate below 10 keV for the smallest W shield. The combination of the energy dependent quenching factor in Ge with the exact neutron energy spectrum results in a count rate comparable to the Pb shield. At higher energies, where the impact of the neutrons is small, an almost identical muon-induced count rate for all shields is deduced. The difference in the atomic number of Pb and W relevant for the *bremsstrahlung* continuum and the self-shielding properties do not lead to a significant discrepancy between the spectra.

### 4.3 Potential background from self-contamination of shield material

The advantages of W over Pb, Cu, Fe and other materials in terms of external $\gamma$-ray and muon-induced neutron background suppression (cf. Sections 4.1 and 4.2) can be diminished by an enhanced self-contamination of the shield material itself. This intrinsic radiopurity is crucial for the decision on how close a material can be deployed to a low background detector.

To start our comparison study, we filled alternately the GIOVE sample chamber with large radiopure samples of W, Pb, Cu and SST and measured their energy spectra with the corresponding HPGe detector. The very low background conditions inside the GIOVE shield allow to study the background that originates prevalently from the four samples. The measured spectra are depicted in Figure 10.
In the case of our 104 kg radiopure W sample, the most important contributions come from the first part of the $^{238}$U chain, namely from $^{234}$Th and its daughter $^{234}$Pa with an activity of $(305\pm30)$ mBq kg$^{-1}$ (cf. Table 2). For the second half of the U decay chain starting with $^{226}$Ra only a small activity of $(4.2\pm0.3)$ mBq kg$^{-1}$ was found; the spectral contributions from $^{214}$Pb and $^{214}$Bi are minor, but visible. The same holds for the $^{232}$Th decay chain, i.e. $(1.0\pm0.2)$ mBq kg$^{-1}$ and for $^{40}$K, i.e. $(5.0\pm0.9)$ mBq kg$^{-1}$.
For the radiopure SST we used the 36 kg 'D6' sample described in Ref. [5]. For the Th, U and K activities only upper limits $<2$ mBq kg$^{-1}$ were derived and no $\gamma$-ray line contributions are visible. The $^{60}$Co activ-ity, which is typically high in SST, is here present at a level of only $(6.1\pm0.9)\,\mathrm{mBq\,kg^{-1}}$. Still, the integral count rate below the $^{60}$Co line at 1.3 MeV is by a factor $\sim$1.5 larger than in the W spectrum.

**Table 4**: Count rates in various energy ranges observed in Ge due to muon-induced secondaries created inside the shield consisting of either 15 or 25 cm Pb or a W shield of comparable shielding power (see Figure 6). The corresponding neutron fluence rates at the diode normalized to the diode surface are given as well. The neutron energies are defined as follows: thermal $[1.0 \times 10^{-9}, 4.0 \times 10^{-7}]$ MeV; intermediate $[4.0 \times 10^{-7}, 0.1]$ MeV; fast $[0.1, 19.6]$ MeV. All results are determined by MC simulations, only statistic uncertainties are given.

| | Ge count rate in $d^{-1}kg^{-1}$ | | | | neutron fluence rate in $cm^{-2}s^{-1}$ | | | |
|---|---|---|---|---|---|---|---|---|
| energy/keV | [1,10] | [10, 100] | [100, 2700] | [40, 2700] | all | thermal | intermediate | fast |
| Pb 15 cm, 1200 kg | 2120±30 | 2040±30 | 17400±100 | 18300±100 | 74.8±0.4 | 0.10±0.01 | 15.6±0.2 | 59.1±0.4 |
| W 9.2 cm, 810 kg | 2050±30 | 1670±30 | 17500±100 | 18400±100 | 83.3±0.4 | 0.06±0.01 | 33.2±0.3 | 50.0±0.3 |
| Pb 25 cm, 3530 kg | 4960±30 | 2700±30 | 17200±100 | 18200±100 | 167.4±0.6 | 0.34±0.03 | 47.4±0.3 | 119.6±0.5 |
| W 15.3 cm, 2150 kg | 3320±30 | 1760±30 | 17300±100 | 18200±100 | 133.0±0.5 | 0.17±0.02 | 63.7±0.4 | 69.1±0.4 |

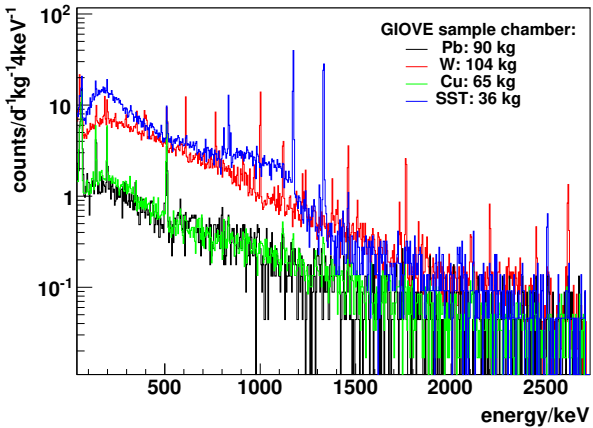

**Figure 10**: Comparison of measured energy spectra from large mostly radiopure samples that were deployed into the sample chamber of the GIOVE screening detector: W, Pb, Cu and Fe-based SST.

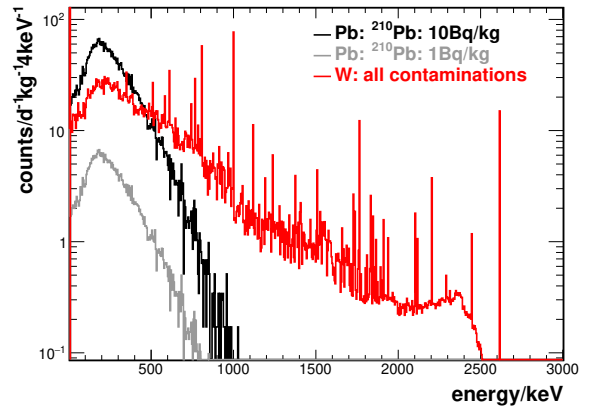

**Figure 11**: Comparison of the impact of the intrinsic contaminations in W and Pb: for W the previously evaluated contaminations from Table 2 are applied, for Pb only a contamination in $^{210}$Pb is assumed. It becomes evident, that with the available W material improvements in the background below 500 keV can already be achieved in comparison to Pb with a $10\,\mathrm{Bq\,kg^{-1}}$ $^{210}$Pb activity.

As radiopure Cu we used the 65 kg electroformed Cu sample described in Ref. [28]. In the case of the $^{238}$U and $^{232}$Th decay chains, $^{40}$K and $^{60}$Co, only threshold activities in the range of $<0.1\,\mathrm{mBq\,kg^{-1}}$ were deduced. Lines from natural and artificial radioactivity are not visible at all in the spectrum. The integral count rate is reduced by a factor of $\sim$4 compared to the one from W.

For the measured radiopure 90 kg Pb sample the integral count rate is reduced by a factor of $\sim$4.5 compared to W and no lines from natural and artificial radioactivity are visible. This is in accordance with the fact, that Pb usually inhibits an excellent radiopurity regarding Th and U except for $^{210}$Pb. For $^{228}$Th, upper limits of $<(0.02\text{-}0.07)\,\mathrm{mBq\,kg^{-1}}$ were found in [11]. The same authors reported a limit of $<(0.03\text{-}0.05)\,\mathrm{mBq\,kg^{-1}}$ for $^{226}$Ra. For $^{238}$U a limit of $<0.02\,\mathrm{mBq\,kg^{-1}}$ was observed in [43]. As anticipated in Section 1, $^{210}$Pb can be troublesome, since it can be present in high concentrations (see e.g. [44, 11, 43]). Pb with $^{210}$Pb activities of 1 to $10\,\mathrm{Bq\,kg^{-1}}$ is commercially available, but expensive. The sample shown here is of high quality in the range of $\sim 0.4\,\mathrm{Bq\,kg^{-1}}$. Smaller activities are usually only observed in ancient Pb that is very rare. For $^{40}$K, there is a broader variety in the radiopurity of Pb, with reported finite values of e.g. $(0.5\pm0.2)\,\mathrm{mBq\,kg^{-1}}$ [11, 44] and upper limits of $<0.2\,\mathrm{mBq\,kg^{-1}}$.

Due to the importance of $^{210}$Pb, we simulated carefully the impact of different $^{210}$Pb activities on the intrinsic background level originating from the shield materials. The first shield is made of purely Pb which is assumed to have exclusively a $^{210}$Pb contamination, either 1 or $10\,\mathrm{Bq\,kg^{-1}}$. The second shield con-sists of purely W, wherein the spectral components of all contaminations were simulated and weighted with the measured activities reported in Table 2. In Figure 11 we show the energy spectra from the W and the Pb shield with the two different scaling factors. For $1\,\mathrm{Bq\,kg^{-1}}$ $^{210}$Pb, the overall background rate and spectrum from the Pb shield lies below the one from W. For $10\,\mathrm{Bq\,kg^{-1}}$ $^{210}$Pb, the W shield spectrum falls below the one from the Pb shield for energies $<500\,\mathrm{keV}$. Equal integral count rates below $500\,\mathrm{keV}$ are achieved for a $^{210}$Pb contamination of 5-6 $\mathrm{Bq\,kg^{-1}}$.

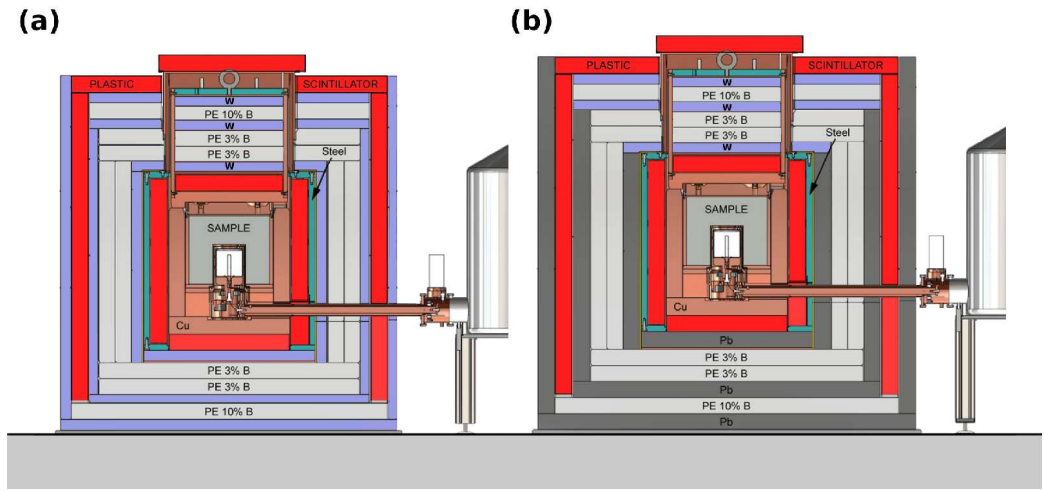

**Figure 12**: Extended shield designs for the GIOVE detector including W: (a) all Pb layers replaced by W, (b) the upper Pb layers are replaced with W.

### 4.4 Simulation of more complex shield designs with Pb and W-Pb combinations

The previously described simple shield in Figure 6 is often extended for most low background applications, depending also on the location (environmental radiation and overburden) and the materials at hand. In the following, we simulate exemplarily modified shields of the existing GIOVE [28] shield setup. The original shield consists of 15 cm Pb in all directions. Compared to the simple shield design in Figure 6 it further features: (1) a muon veto and layers of (borated) polyethylene to moderate and capture muon-induced neutrons, (2) an innermost layer made out of Cu to diminish a potential intrinsic contamination of the shield materials and (3) a chamber with an empty volume of ∼13 l around the detector end cap to install screening samples.

As shown in Figure 12, two shield design modifications by replacing Pb with W in the GIOVE setup were investigated. In the case (a), all Pb layers are replaced with W. The thickness of the other layers and the layer sequence are kept unchanged. As shown in Section 4.1, an almost equal suppression of the external γ-ray background above 100 keV is expected. As a direct consequence the size of the shield is notably reduced and especially the cryostat cooling finger of the HPGe detector can be shortened by about 6 cm. In the case (b), only the top Pb layers are replaced with W to save on costs.

For both shield versions, the muon-induced background was simulated and compared to the original W-free shield design. First of all, it is noted that the inclusion of hydrogenous materials and Cu results in a reduction of the overall neutron fluence rate by a factor of three (cf. Table 5 with Table 4 belonging to the test shields with purely Pb or W). However, the γ-ray background is enhanced because of the self-shielding effect of the *bremsstrahlung*. This is more effective for high-Z materials such as Pb or W than for Cu. The overall neutron fluence rate is diminished by 4% for version (a), with an enhanced reduction observable for fast neutrons. The neutron fluence rate in version (b) is within the statistics uncertainties of the collected simulation output compatible with the neutron fluence rate within the existing GIOVE shield. The electromagnetic background is not significantly lower. In comparison to the test setup in Figure 6, the full impact of the W replacement is not so evident due to the addition of the innermost Cu layer. The neutron fluence rate inside the detector chamber is largely dominated by the innermost shield layer. Neutrons that are created in the outer layers of the shield are already effectively suppressed by the intermediate layers of borated polyethylene. The neutron fluence rate present at the HPGe diode culminates at an energy between the peak energy of Pb and W, as illustrated in Figure 13. However, replacing W with Pb also means that the thickness of the layers of hydrogenous materials can be reduced making the shield even more compact.

**Table 5**: Count rates in various energy ranges observed in Ge due to muon-induced secondaries created inside the GIOVE shield and modifications thereof including W. The modified shields are displayed in Figure 12. The corresponding neutron fluence rates at the diode normalized to the diode surface are given as well. The neutron energies are defined as in Table 4. All results are determined by MC simulations, only statistic uncertainties are given.

| energy/keV | Ge count rate in $d^{-1}kg^{-1}$ | | | | neutron fluence rate in $cm^{-2}s^{-1}$ | | | |
|---|---|---|---|---|---|---|---|---|
| | [1,10] | [10, 100] | [100, 2700] | [40, 2700] | all | thermal | intermediate | fast |
| GIOVE | 980±20 | 2850±30 | 27100±100 | 28900±100 | 29.5±0.2 | 0.26±0.02 | 10.2±0.1 | 19.0±0.2 |
| version (a) | 980±20 | 2830±30 | 27000±100 | 28900±100 | 28.5±0.2 | 0.22±0.02 | 10.0±0.1 | 18.1±0.2 |
| version (b) | 1000±20 | 2850±30 | 27100±100 | 29000±100 | 29.5±0.2 | 0.19±0.02 | 10.3±0.1 | 19.0±0.2 |

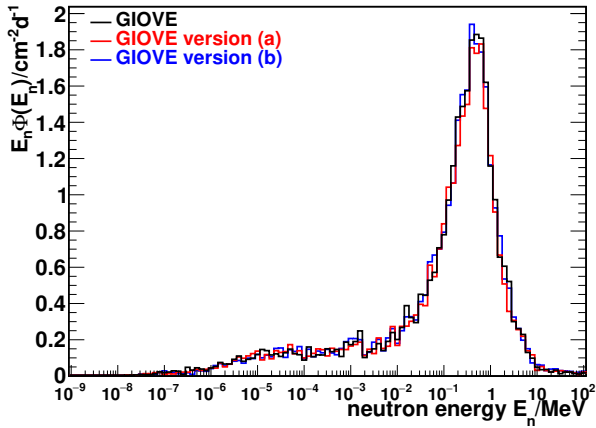

**Figure 13**: MC simulation of the muon-induced neutron fluence rate at the diode surface of the GIOVE detector and two modified shield designs including W.

### 4.5 Realisation of a prototype shield with Pb and W-Pb combinations

A small shield was set up in the MPIK underground laboratory to test the background reduction capability of the new W plates described in Section 3.3. A shield made up of approximately 15 cm of Pb in all directions was compared to a W-Pb hybrid design. This amount of high density materials guaranteed that the measured integral background by the HPGe detector included inside these shield configurations is dominated by muon-induced secondaries.

For the W-Pb hybrid design, 28 of the described W plates were used to fully replace the Pb shield above the end cap and partially the innermost vertical Pb layers around the copper end cap (cf. photos in Figure 14). The mass of the shield is reduced by ∼200 kg in this way, while providing approximately the same shield capability against external $\gamma$-radiation. Moreover, a plastic scintillator plate equipped with one PMT is available and operated as a real-time muon veto. It is placed on top of the shield and leads to a suppression efficiency of about 60%.

The results are summarized in Table 6 and are plotted in Figure 15. The spectra with both shield configurations are comparable above 500 keV as expected because here the electromagnetic component dominates. Below 500 keV, including W into the shield leads to a reduction in the observed background. Both, the $^{210}$Pb activity of the Pb – in this case not especially radiopure – and the lower fast neutron fluence rate contribute to the overall better performance of the hybrid shield. The $^{210}$Pb activity is more relevant above 100 keV, while below the impact of the reduced neutron fluence rate increases towards lower energies. The reduction of the neutron fluence rate is confirmed via an independent study of characteristic $\gamma$-ray lines induced by fast and thermal neutrons. They can be used to predict and compare the respective fluence rates at the diode. Fast neutrons can scatter inelastically in Ge. The $\gamma$-rays emitted after the deexcitation of the hit nucleus add up with the recoil of the nucleus resulting in the typical 'saw-tooth' structure of the peaks. In the collected spectra, such kind of peak corresponding to $^{74}$Ge(n,n'$\gamma$) at 596 keV is observable. In the determination of the count rate, the 609 keV line from $^{214}$Bi (determined from background data with applied muon veto) has to be subtracted. In the hybrid shield design, the count rate within the line is lower by (25±10)%, confirming the reduction in the fast neutron fluence rate. Additionally, the 139.5 keV line created in neutron capture on $^{74}$Ge was studied. The line is only visible, if the muon veto is applied. The neutron capture product $^{75m}$Ge itself is metastable (T$_{1/2}$=47.7 s) with a life-time longer than the veto window of 160 $\mu$s. Thus, it becomes evident after the application of the muon veto. A comparable count rate for both shield version, i.e. (16±1) counts d$^{-1}$, was determined. This means that no significant difference in the fluence rate of neutrons of low energies was observed.

All in all, our prototype W-Pb hybrid shield con-firmed the findings of the previous sections.

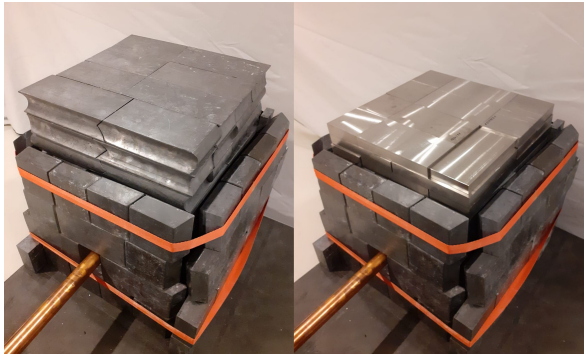

**Figure 14**: Photos of the prototype shields including the CONRAD HPGe detector: On the left, a shield design consisting of 15 cm Pb in all directions is shown. On the right, the top layer and part of the innermost layer were replaced by an equivalent amount of W pseudo-alloy plates reducing the total shield mass by about 200 kg. For visibility purposes, the muon veto was removed from the top.

## 5  Summary and Conclusions

Within this work an extensive search for radiopure high density W for low background applications has been successfully conducted. For the first time, one manufacturer was singled out that can produced W plates with contaminations of only $1\,\mathrm{mBq\,kg^{-1}}$ level for most natural and artificial radionuclides. Only the first half of the $^{238}$U decay chain was found to be elevated at the order of $0.8\,\mathrm{Bq\,kg^{-1}}$. A careful study and selection of powders and additives entering the W pseudo-alloy production allowed for a significant reduction of this contaminant by a factor of ∼2.5. Surface treatments via mechanical/chemical cleaning and baking did not further reduce the $^{238}$U contamination pointing towards a persisting bulk contamination. Further options to create W pseudo-alloys with an even higher level of radiopurity exists (e.g. cleaning of W powders, replacement of clay with ceramics as new substrate, sintering under $N_2$ atmosphere), however they still need to be investigated. Nonetheless, we were able to proof that the production chain and the obtained radiopurity levels are reproducible.

Besides the natural and artificial contaminants we investigated the impact of cosmic radiation on the total radioactivity budget of W. The fast hadronic component of cosmic rays at Earth's surface can create long-lived radioactive isotopes. Specifically, we identified $^{182}_{73}$Ta, which might be problematic due to its half-life of 114.6 d and the emission of $\gamma$-rays of 1814 keV in connection with a measured sea level production rate of $(121\pm30)\,\mathrm{atoms\,d^{-1}}$. Long-lived iso-topes can also be activated by neutron capture of predominately thermal neutrons ($^{187}$W and $^{187}$W). Next to the hadronic component of the cosmic rays, thermal neutrons can also be created *in situ* when muon-induced neutrons in the shield are moderated. No significant *in situ* activation of $^{187}$W and $^{187}$W has been observed in our material screening data. Further, we investigated prompt and delayed signals from muon-induced secondaries that are created inside the high density layers of a shield made of Pb or W. This is of special interest at shallow depth locations, where the cosmic ray muon flux is not suppressed very well by the overburden. While the electromagnetic *bremsstrahlung* continuum extends over the full spectral range, the elastic and inelastic scattering of neutrons in a detector leads to an enhancement of the continuum at low energies by energy depositions of recoiling target nuclei.

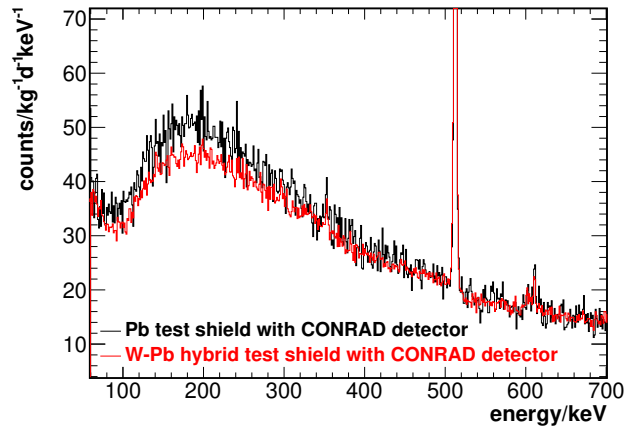

**Figure 15**: Experimental comparison of a pure Pb shield of 15 cm to a W-Pb hybrid shield, where the layers above the detector have been replaces with the equivalent amount of W. All data are collected with the CONRAD detector. The reduced neutron fluence rate becomes evident due to the reduced continuum at low energies as well as the suppressed line at 596 keV from inelastic neutron scattering on Ge. Additionally, the smaller activity of $^{210}$Pb contributes to a reduction of the continuum in the W-Pb hybrid shield.

Many low background applications based on radiopure high density W shields are imaginable, however the benefit depends on the energy regions of interest, the expected background types, the overburden of the locations, as well as other boundary conditions. Especially for applications in the region below a few 100 keV, where $^{210}$Pb $\beta$-decays and neutron-induced nuclear recoils play a major role, W-based shields become advantageous compared to radiopure Pb shields. This includes a wide range of applications, from dark matter searches at deep underground sites to neutrino investigation programs at shallow depth close to nuclear power plants or spallation neutron sources. For example, the signal expected from a neutrino magnetic moment is supposed to increase approximately like one over energy towards low energies and thus such applications can benefit from a radiopure W shield. Next to shields for low background applications, radiopure high density W might also be of interest in a thermal dissociator where a hot W surface (∼2500 K) is used to crack molecular tritium to produce atomic tritium. This is, for instance, planned for experiments such as Project 8 aiming at the detection of the absolute neutrino mass [45]. Any additional electron background might diminish the sensitivity of endpoint measurement of the tritium $\beta$-decay. Beyond fundamental physics, radiopurity is also relevant in nuclear medicine, where high density W is used as radiation shield close to human beings [15].

Table 6: To compare the W-Pb hybrid design to the pure Pb shield, integral count rates in various energy ranges as well as the count rate in a line created in fast neutron inelastic scattering at 595 keV and in neutron capture at 139.5 keV are evaluated. The count rates marked with (*) are derived from data with applied muon veto. The hybrid shield reduces the overall mass of the setup by ∼200 kg.

| shield | meas. time/d | [60,100] keV | [100,500] keV | [500,2700] keV | 596 keV line | 139.5 keV line |
|---|---|---|---|---|---|---|
| Pb, ∼1200 kg | 2.9 | 2700±30 | 27100±100 | 25400±100 | 109±9 | 16±1 (*) |
| W-Pb hybrid, ∼1000 kg | 11.2 | 2500±20 | 25100±50 | 25300±50 | 81±5 | 16±1 (*) |

In the course of our study, we investigated exemplarily a few shield configurations via validated MC simulations for the shallow depth case with an overburden of 15 m w.e. and performed measurements with a HPGe detector inside a prototype W-Pb hybrid setup. First of all, we compared pure Pb and W shields. Replacing a 15 cm thick Pb (minimum required to suppress natural radiation) by 9.2 cm of W, the equivalent shield thickness in the suppression capability of $\gamma$-radiation, results in a significant reduction of the fast neutron fluence rate at the detector by 17%. This goes along with a diminished background count rate below 100 keV. The effect is more pronounced for larger shields, meaning that in an environment with a large $\gamma$-ray background e.g. close to a reactor core of a nuclear power plant, a W shield is even more beneficial compared to a Pb shield. Including hydrogenous materials and an inner layer of Cu into the shield, such as for the material screening detector GIOVE, already itself strongly reduces the neutron fluence rate at the HPGe diode. Nevertheless, replacing Pb by W in the design still leads to a lower neutron fluence rate by about 4% and the advantages of a more compact shield. If no inner layer of Cu is present in the shield design (which results in a lower *bremsstrahlung* continuum, the expected gain from an interchange of Pb and W is expected to be even higher by more than 10%.

Beside the potential gain in background suppression, W-based shields offer several more advantages compared to Pb. Next to safety aspects (health care during handling, fire-load reduction, mechanical stability) the higher density allows to build more compact shields, as demonstrated in our examples. This compactness has several benefits. It is economically of interest: even though the identified high density W composite is similarly expensive to Pb with a $^{210}$Pb activity of 1 Bq kg$^{-1}$, other cost-intensive shield parts like (borated) polyethylene layers and muon veto systems can be reduced in size. Further, experimental sites such as locations at a nuclear power plant might have spatial limitations to accommodate a full setup. Next to it, the total mass of the setup can be minimized helping in critical situations imposed by structural engineering calculations. Finally, the shielded detectors themselves profit as well, since a smaller shield implies e.g. more compact cryostat systems, thus better cooling conductance and – as an ultimate consequence – a better detector performance in terms of energy threshold and resolution for HPGe spectrometers.

To conclude, there are other interesting high density materials next to W, that can be explored under similar considerations in the near future. Comparable to the described W pseudo-alloy with Fe and Ni additives, also W composites with a higher Cu content (10% Cu; $\rho$=16.75 g cm$^{-3}$) or carbon additives ($\rho$=15.63 g cm$^{-3}$) are available on the market. Even though their densities are reduced, the differences in the manufacturing process might turn into an improved radiopurity level. Other potential candidates beyond W are gold (Au) ($\rho$= 19.32 g cm$^{-3}$) and Ta ($\rho$=16.65 g cm$^{-3}$). While Au is supposed to be radiopure but beyond the scope of any practical use, Ta might be another promising candidate for low background applications.

## 6 Acknowledgements

produced by different manufacturers for the experimental stellarator facility Wendelstein 7-X. We are deeply in debt with B. Döbling and the production team from H.C. Starck Hermsdorf GmbH in Hermsdorf, Germany, for a highly prolific and enduring support of this project over several years.

For technical support we thank the mechanical workshop at the Max-Planck-Institut für Kernphysik in Heidelberg, Germany. We kindly thank H. Strecker and Dr. G. Heusser for assistance during the tungsten measurement campaign, and Dr. M. Laubenstein for sharing with us previously obtained screening results from high density tungsten samples.

This work was supported financially by the Max Planck Society (MPG).